# Feature-Resolved Photoluminescence Analysis: Probing Emission Beyond Conventional Photon Statistics

Amit R. Dhawan[1,2*], Nishita Chowdhury[1], Willy D. de Marcillac[1], Michel Nasilowski[3], Benoît Dubertret[3], Agnès Maître[1*]

[1]Sorbonne Université, CNRS, Institut des Nanosciences de Paris, INSP, F-75005 Paris, France
[2]Department of Physics, University of Oxford, Oxford OX1 3PU, United Kingdom
[3]Laboratoire de Physique et d'Etude des Matériaux, ESPCI-ParisTech, PSL Research University, Sorbonne University, CNRS UMR 8213, 10 rue Vauquelin, Paris 75005, France

*Corresponding authors.
Email(s): agnes.maitre@insp.upmc.fr, amit.dhawan@physics.ox.ac.uk

We present a feature-resolved methodology to analyse the photoluminescence dynamics of single emitters using a combination of lifetime, spectral, and photon correlation analyses. By integrating conventional ensemble photon statistics measurements with emission state-resolved, spectrally filtered, and lifetime-gated methods, we uncover emission dynamics that remain hidden in ensemble treatment. We study the fluorescence of single CdSe/CdS core/shell colloidal quantum dots under varying excitation powers. Using feature-resolved analysis, we understand the radiative and non-radiative recombination processes, and estimate quantum parameters. Event-selective analysis provides a versatile toolkit for characterizing emitters, both single and aggregate particles. These methods are broadly applicable to a wide class of photoluminescent emitters, such as nitrogen vacancy centres in nanodiamond, epitaxial quantum dots, and perovskite nanocrystals. The application of these lateral investigation techniques will contribute to the advancement of quantum light source development.

The 21st century is witnessing a photonic technology revolution, and the success of several of these technologies is tied to the performance of their photon sources, which has advanced research aimed at understanding and engineering their photophysical properties. Colloidal quantum dots (QDs), also known as core-shell nanocrystals, are promising light sources that are widely studied and used for several applications such as light emitting devices,[1–3] photovoltaics,[4–7] biomedical,[8,9] and quantum systems.[10–13] QD research has seen significant development since Ekimov produced QDs in glass in 1981, Brus's synthesis of colloidal QDs in 1984,[14] the first report on QD fluorescence blinking in 1996,[15] attempts towards creating "giant" non-blinking QDs in late 2000s,[16,17] to the exploration of alternative QD structures such as gradient QDs,[18] to tuneable lasing with QDs,[19] to their use in commercial applications today.

The photophysical properties of chemically synthesised colloidal QDs can be customtuned by modifying their dimension or composition. It may involve new and interesting photo-physical phenomena that open doors to further investigation and applications. Their absorption and emission can be modified intrinsically or extrinsically. The former includes controlling the materials, size and arrangement of the QD core and shell, and the latter involves changing the environment of the QDs, which can be the surface chemistry of the QD[20–22] or the system that embeds it.[23–25] While there are valid use cases for macroscopic or large-scale study of QDs in dispersions, films and aggregates, understanding QD photo-physics[26] at a quantum level necessitates single-emitter analysis. Moreover, analysis of certain phenomena, such as single-photon emission and multi-exciton decay, requires experimenting with only one QD at a time.

The thick shell in giant CdSe/CdS core/shell quantum dots permits radiative multi-exciton recombination,[18] which increases significantly with excitation energy.[23] It makes them almost non-blinking, bright, and suitable for many applications. Single-photon emitter giant QDs are desirable, and researchers have made progress in their fabrication.[27] A detailed understanding of their emission decay and photodegradation mechanisms is crucial for enhancing and adapting them for specific applications. These insights can be used to advance fabrication technology that will address existing limitations and create more efficient QDs. It is also commercially interesting as colloidal QDs are used in consumer and industrial products.

Here, we present feature-resolved techniques to analyse the emission statistics (lifetime and photon correlation) and spectra of fluorescent single emitters, and extract parameters such as absorption cross-section, photoluminescence quantum efficiency, and energy transitions. We show that integrating standard photon statistics measurements (overall lifetime and photon correlation) with emission state-resolved, spectrum-resolved, and lifetime-gated correlation analyses enables the understanding of the photo-physical processes that would otherwise remain concealed. These methods, demonstrated on single CdSe/CdS giant QDs, are broadly applicable to most photoluminescent single emitters. Examples include nitrogen vacancy centres in diamond[28], perovskite[29] and epitaxial[30] QDs, which are relevant to photonic quantum technologies.

## Results

The absorption of light of appropriate wavelength by a colloidal QD creates charge carriers in it, which recombine to emit photoluminescence. The recombination of charge carrier in a QD occurs via radiative and non-radiative channels, and can include several mechanisms, such as excitonic, multi-excitonic, trions, and Auger processes. The emission from colloidal QDs is affected by the Auger effect, where the energy released from an electron-hole recombination is used to excite another charge carrier and is not emitted as a photon, thus making the transition non-radiative. Under weak pumping or in small thin-shell QDs, most multi-excitonic radiation is suppressed by non-radiative Auger effect, and the final excitonic recombination yields single photons. As the Auger recombination is inversely proportional to the QD volume,[31,32] it is significantly reduced in giant QDs, which can make their emission muti-excitonic. This affects their photon throughput, lifetime, and photon-correlation. Under low pumping, a typical QD from the tested batch emitted single photons with 70 − 90% purity. Generally, as the pump fluence increases over a QD that has not degraded, its lifetime decreases, and its photon-emission rate and $g^{(2)}(0)$ increase due to multi-excitonic recombination. In addition to the CdSe/CdS core-shell emission, there can be bulk-like shell emission, which has remarkably distinct spectral and temporal signatures.

Here we analyse our measurements on individual CdSe/CdS core/shell giant QDs (core diameter 3 nm and shell-thickness 5–8 nm) excited by a 405 nm pulsed laser (100 ps pulse width, 2.5 MHz repetition rate, see **Methods**) focused through a 0.8NA air objective. The QDs were spin-coated on a 0.17 mm thick glass slide.

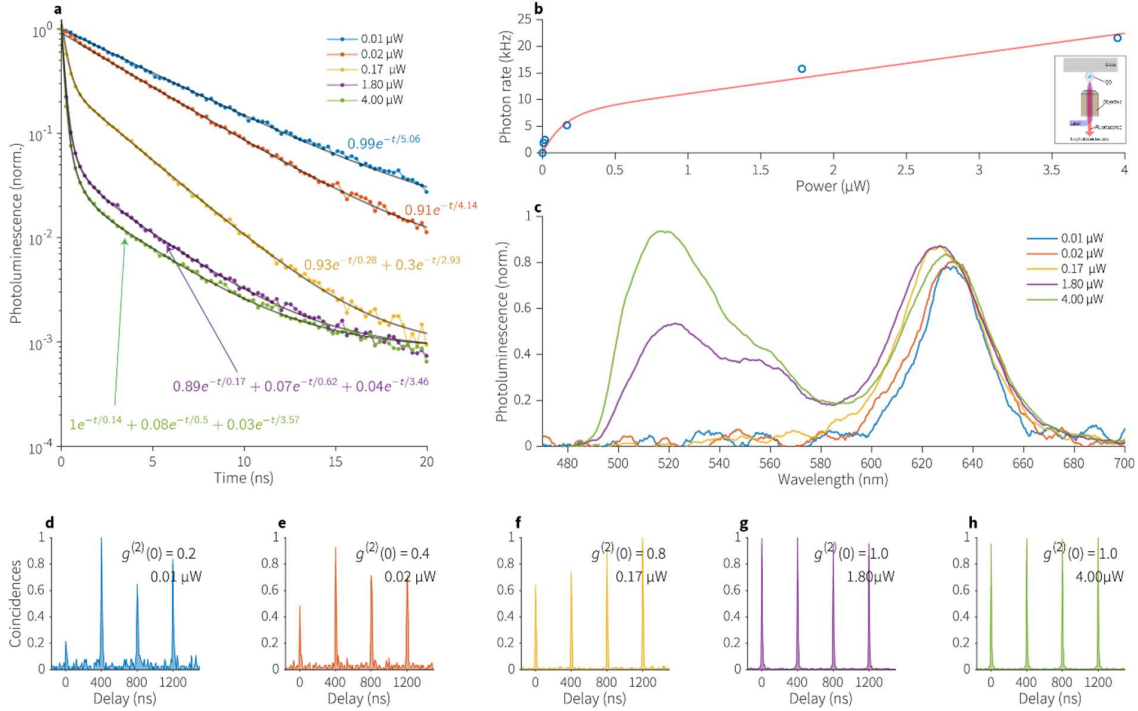

**Figure 1**. QD photoluminescence decay with fitting functions in corresponding colours (a), emission rate (b), spectra (c), and photon correlation (d–h) measured at five power levels. The increase in excitation power enhanced multi-exciton recombination that reduces the lifetime and increased $g^{(2)}(0)$. It blue-shifted the emission (c), and at $P = 1.8$ μW and 4 μW, $g^{(2)}(0) = 1.0$ (g,h), we note the appearance of another emission wavelength due to bulk-like CdS emission, which is separated by about 120 nm from the typical CdSe/CdS emission. The inset in (b) is the excitation and collection schematic.

## Giant QD emission

Figure 1 shows the response of a QD under increasing laser power $P$. At $P = 0.01$ μW and $0.02$ μW, the mono-exponential decay curves (blue and red curves in Figure 1a) demonstrate that the QD emission was primarily excitonic. The emission remained single-photon with $g^{(2)}(0) < 0.4$ (Figures 1d,e), and its spectra were centred around 635 nm (blue and red curves in Figure 1c). As the excitation power increased from 0.01 μW to 0.02 μW, the lifetime reduced and $g^{(2)}(0)$ increased, which indicates a larger contribution of non-radiative recombination at $P = 0.02$ μW. The total decay rate $\Gamma = \Gamma^r + \Gamma^{nr}$, where $\Gamma^r$, $\Gamma^{nr}$ are the radiative and non-radiative components, respectively. Under weak pulsed laser excitation, $g^{(2)}(0) \approx \eta_{BX}/\eta_X$[33] where $\eta_{BX}$ and $\eta_X$ are the bi-exciton and exciton quantum yields, respectively. Following the definition of quantum yield $\left(\eta = \frac{\Gamma^r}{\Gamma} = \frac{\Gamma^r}{\Gamma^r + \Gamma^{nr}}\right)$ and assuming that $\eta_{BX}$ did not change with power as the decay curve remained mono-exponential, the increase in $g^{(2)}(0)$ at 0.02 μW was due to the decrease in $\eta_X$. In other words, higher excitation energy reduced $\eta_X$.

The photon rate (Figure 1b) increased as the laser power was increased to power to 0.17 µW, and the emission comprised of fast and slow decay components (yellow trace, Figure 1a), which reveals two significant and temporally distinct recombination mechanisms. To understand the nature of these processes that result in a bi-exponential decay curve, the data were processed using temporal filtering. By filtering out the fast lifetime component, we can analyse the slow lifetime events. Removing the photon events in the initial 2.5 ns (Figure 2a) reduced $g^{(2)}(0)$ from 0.8 to 0.3 (Figure 2b,c), thus verifying that the fast component was due to bi-excitonic emission. This was further substantiated by a 10 nm blue shift in the emission spectra (yellow curve, Figure 1c), which is attributed to bi-excitonic emission[34,35]. So far, at $P = 0 - 0.17$ µW, the QD can be modelled as a two-level system, which is at most bi-excitonic.

At $P > 0.2$ µW, the emission spectrum segregated into two distinct peaks (purple and green curves in Figure 1c). The low-energy photoluminescence between $590 - 690$ nm was mainly due to excitonic and bi-excitonic contributions from the CdSe/CdS QD, and the high-energy $490 - 580$ nm photoluminescence was due to bulk-like CdS shell emission[36,37]. At $P = 1.8$ µW and $4$ µW, the emission was significantly fast and multi-excitonic that could not be fit with a bi-exponential function (purple and green traces in Figure 1a), was multi-photonic with $g^{(2)}(0) = 1$ (Figures 1g,h), and it increased without saturating (Figure 1b), which made the two-level system approximation invalid. Removing the photon events in the initial 2.5 ns at 1.80 µW (Figure 2a), decreased $g^{(2)}(0)$ from 1.0 to 0.6 (Figure 2d,e). The reduction in $g^{(2)}(0)$ at 1.80 µW was less compared to 0.17 µW due to a larger contribution of multi-excitonic emission in the former.

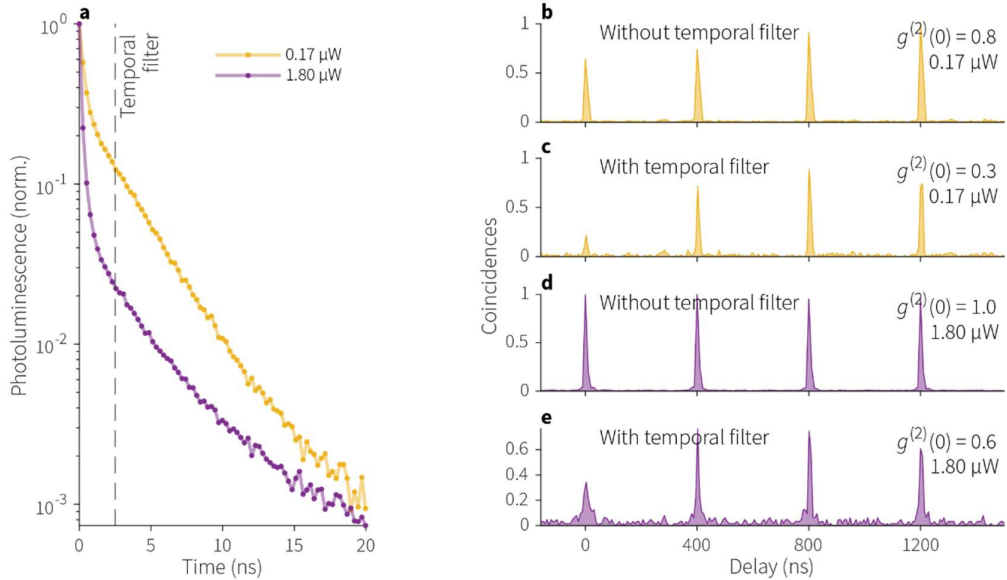

**Figure 2**. QD photon emission events with photoluminescence decay curves (a) at two powers with single photon emission statistics (with and without the temporal filter of 2.5 ns) in corresponding colours (b-e).

*Low-energy and high-energy emission*

We can correlate the information in Figure 1 to understand the emission process and estimate the absorption cross-section and quantum efficiencies. As demonstrated by its spectra, the emission of this QD consisted of two main emission processes, which are modelled in Figure 3a. The typical CdSe/CdS emission occurred under low to mid-level excitation fluence, whereas fast bulk-like CdS emission appeared only at higher excitation fluence. We model the complete emission process as a combination of low energy CdSe/CdS emission two level system, and a high-energy CdS emission system.

The creation of charge carriers in the QD by a laser pulse in both these processes is random and follows Poissonian statistics. The probability of absorbing $k$ laser photons $\mathcal{P}(k;\bar{n}) = \frac{\bar{n}}{k!}e^{-\bar{n}}$, where the mean photon number in one laser pulse $\bar{n} = \sigma_{\text{abs}} \frac{1}{A_{\text{spot}}\left(\frac{hc}{\lambda}\right)f_{\text{rep}}} P$. Here, $\sigma_{\text{abs}}$ is the absorption cross-section of the QD, $h$ is Planck's constant, $c$ is the speed of light in vacuum, and $A_{\text{spot}}, \lambda, f_{\text{rep}}$ and $P$ are the focused spot area, excitation wavelength, repetition rate and averaged power of the laser. Each absorbed photon creates an electron-hole pair in the QD. Figure 3 details the fitting of the photon rate curve of Figure 1b.

We can consider the cascading low-energy CdSe/CdS core-shell transitions at $P = 0 - 0.17$ µW as at most primarily bi-excitonic, which is substantiated by the emission decay curves (Figure 1a). In this two-level system, the probability of creating more than one exciton $\mathcal{P}(>1;\bar{n}) = 1 - \mathcal{P}(0;\bar{n}) - \mathcal{P}(1;\bar{n})$, where $\mathcal{P}(0;\bar{n})$ and $\mathcal{P}(1;\bar{n})$ are the probabilities of creating 0 and 1 exciton, respectively. In these giant QDs, the exciton quantum yield $\eta_X \approx 1$[38] and using the $g^{(2)}(0) = 0.21 \approx \eta_{BX}/\eta_X$ under weak pumping (Figure 1d), we find the bi-exciton quantum yield $\eta_{BX} \approx 0.21$. The total detected photon rate due to the CdSe/CdS core-shell emission (red gradient area in Figures 3b,c)), $\phi_1 = \eta_C f_{\text{rep}}[\eta_X \mathcal{P}(1;\bar{n}) + (\eta_X + \eta_{BX})\mathcal{P}(>1;\bar{n})]$, where the collection efficiency $\eta_C$ includes all opto-electronic system losses, such as light transmission losses of the optical system and the photon detection performance of the single-photon detectors. Our setup directed 70% of fluorescent light to the spectrum analyser; this accounted in $\eta_C$. In the absence of high-energy CdS shell emission (blue gradient area in Figures 3b, c), the photon rate curve of this QD would have followed the dashed curve in Figure 3d as it would saturate in a two-level system. The ratio of low-energy transitions to total transitions was found using the area under its spectrum ($= A_{\text{red}}/[A_{\text{red}} + A_{\text{blue}}]$). The dashed curve in Figure 3d was obtained by multiplying the photon rate at $P = 1.80$ µW and $4.00$ µW with the contribution of only the low-energy CdSe/CdS emission (red gradient areas in Figures 3b,c), which reduced the photon rate at high excitation power. Using the adjusted photon output, we fitted it (dashed curve) with $\phi_1$, and found $\sigma_{\text{abs}} = 2.3 \times 10^{-14}\text{cm}^2$ and $\eta_C = 0.01$ or 1%. The high value of $\sigma_{\text{abs}}$ demonstrates the high quality of these single emitters, and the low value of $\eta_C$ is due to ~10% of fluorescence reaching the air objective; the rest is transmitted into the glass substrate, the higher refractive index medium. Only 1% of the 10% fluorescence is detected by the photodetectors due to the transmission loss of the objective, reflection losses of the mirrors, and 40% photon detection efficiency of the detectors.

For bulk-like CdS emission, we consider three or more photon absorption events. The initial two absorbed photons saturate the low-energy CdSe/CdS transitions. The absorption of three or more photons (defined as $m$ below) results excites high-energy CdS states. The photon rate due to these transitions is given as

$\phi_2 = \eta_C f_{rep}[\sum_{m=3}^{\infty} \mathcal{P}(m; \bar{n})(m-2)\eta_{H_m}]$, where $\eta_{H_m}$ is the quantum efficiency of the recombination related to $m$ electron-hole pairs. In calculation, the infinite number of terms in the sum is limited to $T$, which is sufficiently large to include all transitions. It was 50 in this case. We assumed a constant $\eta_{H_m}$, which was found to be 0.07 by fitting the total photon rate with $\phi_1 + \phi_2$ (red curve in Figure 3d). With respect to the fit, the lower power data point is slightly above it, whereas the higher power photon rate is below it. This suggests that $\eta_{H_m}$ reduced as the excitation power increased. Indeed, higher excitation creates more excitons, thereby increasing the contribution of non-radiative Auger processes. Although multi-exciton quantum efficiencies in typical CdSe/CdS transitions have been investigated,[39] bulk-like CdS emission from CdSe/CdS QDs lacks similar research. The expression of $\phi_2$ can be modified to include quantum efficiency scaling information, if available.

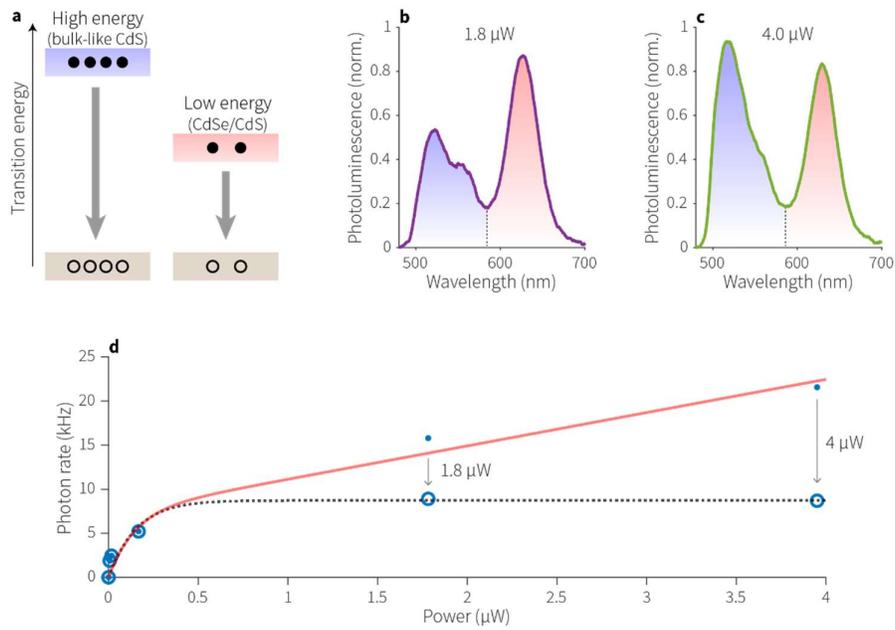

**Figure 3.** (a) Energy level illustration depicting low-energy (red gradient wavelengths in (b,c)) and high-energy (blue gradient wavelengths in (b.c)). (b, c) QD emission spectra at two powers with showing CdSe/CdS emission (red gradient) and bulk-like CdS emission (blue gradient), separated by a dashed line. (d) QD photon rate with incident power (red curve), and adjusted photon rate after removing the bulk-like CdS contribution (dashed curve).

## QD emission states

QD emission is generally composed different emission states including dark, grey, bright states, which can be resolved, and measured separately for lifetime and photon correlation. These measurements yield information that cannot be obtained through conventional analysis. In a batch of QDs, some show atypical behaviour that provides insight into the relaxation dynamics. Atypical emission could be due to an inherent property of the QD or a sign of degradation, such as photo-oxidation. If the emission characteristics of a QD at a given excitation power do not change significantly during the experiment, it confirms that the QD has not degraded due to excitation.

Here we report measurements on a non-degraded QD. The plots in Figure 4 were obtained by exciting a QD at different excitation intensities using a 1.4NA oil objective. In this case, however, as the excitation over the QD was increased, its lifetime (Figure 4a) did not always decrease at higher excitation, and the emission rate and $g^{(2)}(0)$ did not increase monotonically with excitation (Figure 4b). The data points in Figure 4b are connected through interpolation.

The gradient-shaded regions in Figure 4b highlight the demarcation between the usual (R1, green overlay), unusual (R2, red overlay), and saturating (R3, purple overlay) responses. The response was typical of a quasi-two-level system in R1 at the first three measurement points of $P = 0.02, 0.03,$ and $0.04\ \mu W$ (in R1), where higher excitation linearly increased the photon rate. The lifetime at $0.02\ \mu W$ and $0.03\ \mu W$ remained similar, and it slightly decreased at $0.04\ \mu W$. In all three cases, $g^{(2)}(0) \approx 0.1$.

In R2, under increasing excitation, the photon rate decreased and $g^{(2)}(0)$ increased. This was due to decrease in $\eta_X$ caused by non-radiative excitonic recombination and intermittent trap charges. The latter is substantiated by the blinking behaviour of the QD in the following text.

In region R3, at $P = 0.5 - 1.5\ \mu W$, the photon rate increased and showed saturation at higher powers, and $g^{(2)}(0)$ remained high throughout. The high $g^{(2)}(0)$ is attributed to increased multi-excitonic recombination under higher fluence, which is evident from the short lifetime at these powers. At $P > 1.5\ \mu W$, the photon rate decreased due to non-radiative effects being more effective than radiative multi-excitonic recombination.

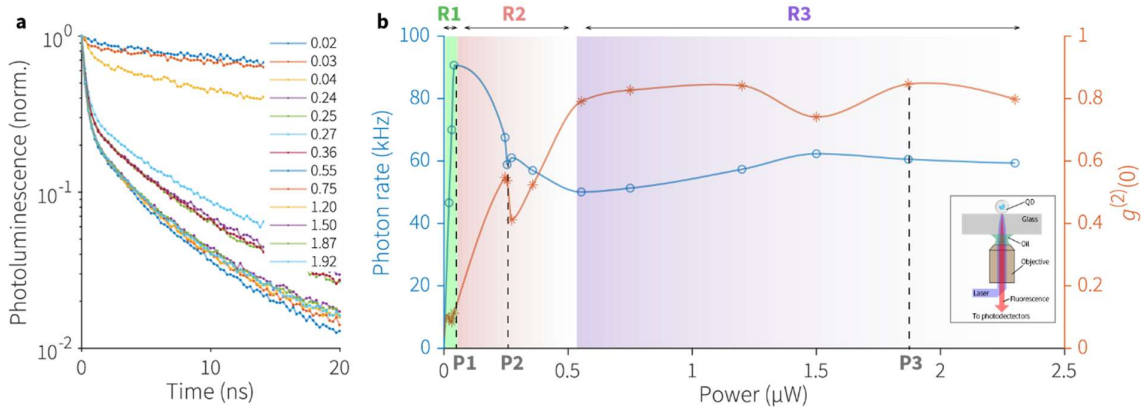

**Figure 4**. a) QD decay under increasing excitation fluence. The legend entries in (a) are the excitation power readings in µW corresponding to the measurement points in (b), which plots the photon detection rate and $g^{(2)}(0)$ as a function of excitation power. Region R1 shaded in green is usual QD response, red-shaded region highlights the departure from the nominal, and purple-shaded region shows normal behaviour. The three power levels P1, P2, and P3 are 0.04 µW, 0.23 µW, and 1.87 µW, resp. Further characterisation of emission at these powers is in Figure 4. The inset in (b) is the excitation and collection schematic.

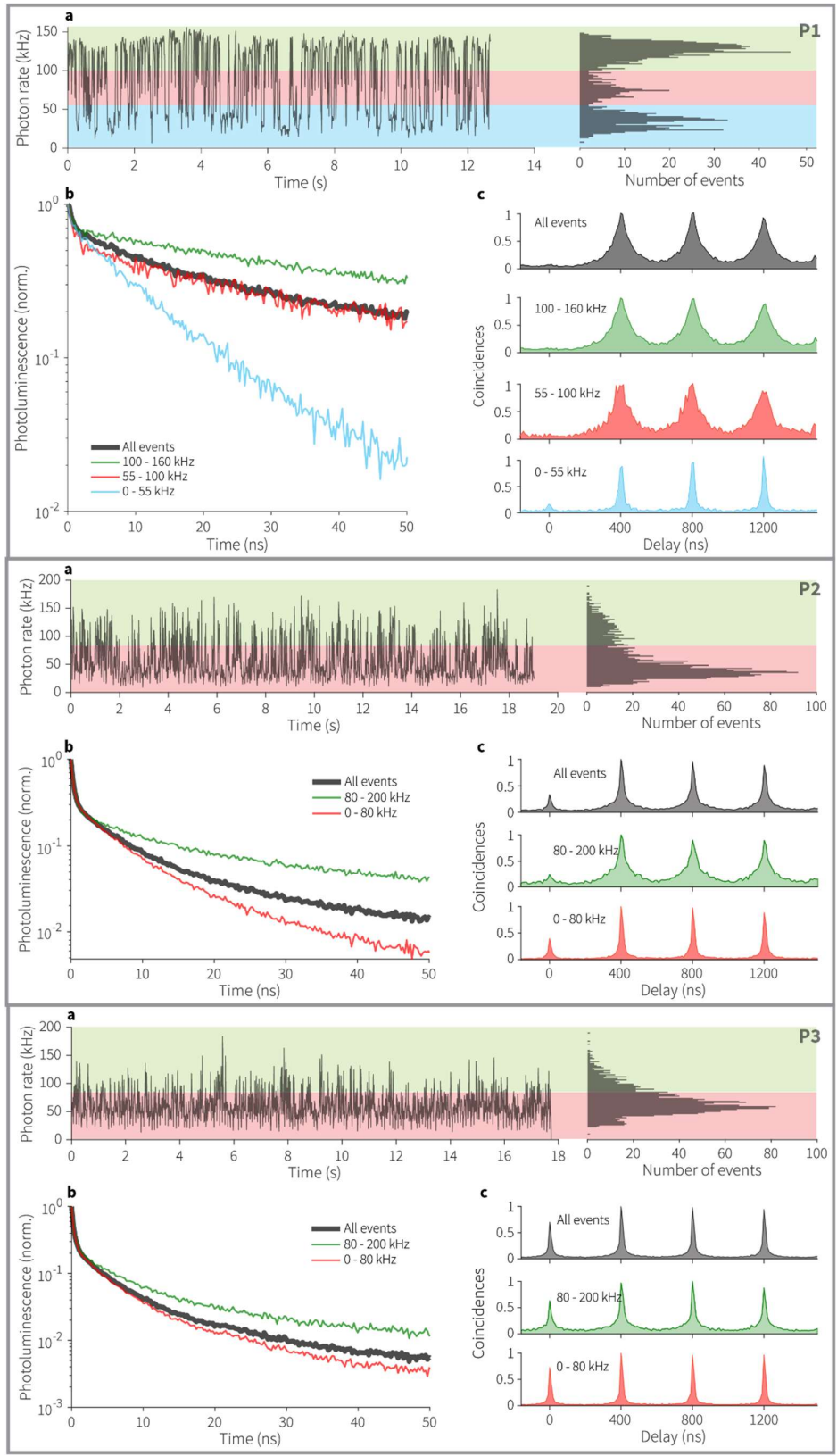

**Figure 5**. The boxed measurements were obtained at power levels P1, P2, and P3 (0.04 µW, 0.23 µW, and 1.87 µW), resp. as detailed in Figure 3. At a given power, (a) depicts the photon detection rate and its histogram (bin width = 0.01 s)

highlighting prominent emission states. The lifetime (b) and photon correlation (b) of the emission states are plotted include the photon rate range.

The P1 box in Figure 5 shows that the QD emission comprised of three prominent emission states (a), and their lifetimes and photon correlation in (b) and (c), resp. The overall emission lifetime (black curve), which had contributions from all the emission states, was faster than the bright emission state (green curve) and slower than the dark state (cyan curve), and similar to the grey state (red curve). In this weak pumping, the higher single-photon purity of the bright state is substantiated by its background noise limited $g^{(2)}(0)$ (green curve in (c)), which is almost zero due to its almost negligible $\eta_{BX}$. It follows $g^{(2)}(0) \approx \eta_{BX}/\eta_X$, with $\eta_X \approx 1$.[40] The $g^{(2)}(0)$ peak of the dark state emission (cyan curve in (c)) was higher than in other states due to higher non-radiative recombination that reduces $\eta_X$, and therefore increases $g^{(2)}(0)$. The low photon generation rate of the dark and grey states suggests $\eta_X < 1$ and the presence of non-radiative recombination. Here non-radiative recombination is faster than radiative recombination, which made these states quicker than bright state emission. The initial rapid component in all lifetime curves was due to bi-excitonic recombination, which did not have a major contribution to the overall emission.

A QD emits usually emits more photons under higher fluence. The photon output of this QD, however, reduced at $P > 0.1\ \mu W$. The P2 box of Figure 5 shows a prominent dark state (a) with a $g^{(2)}(0) = 0.5$ (red curve in (c)). Here as well, the brighter emission state was slower than the dark state. However, the difference was comparatively less than in P1. The contribution of the dark state (pink overlay in inset(a)) was significantly higher than the bright state (green overlay), which can be attributed to abundant charge trapping (trions).

The inset (a) of P3 measurements depicts only one prominent emission state, which was faster than the bright emission (b). All emission states showed $g^{(2)}(0) > 0.5$, which indicates significant non-radiative recombination ($\eta_X < 1$) or more multi-exciton recombination. The strong presence of dark state signifies non-radiative recombination and an overall $g^{(2)}(0) \approx 0.8$ indicates strong multi-exciton recombination.

## Conclusion

We detailed aspects of QD emission and showed that feature-resolved post-processing of photon events can uncover relaxation dynamics that cannot be accessed through conventional analysis. These methods be extended to other quantum emitters. Moreover, event-specific photon data processing can be implemented in real-time data analysis. This will reduce the excitation duration, and it will be particularly useful in investigating novel classes of emitters that have limited photon throughput. Integrating advanced analysis tools into commercial photon counting systems will be well received by engineers and scientists, as it enables new avenues of research without incurring additional costs.

## Methods

**Sample preparation:** The CdSe/CdS core/shell QDs were synthesized chemically using the slow injection method[41,42]. QDs in hexane were spin-coated at 4000 rpm for 40 s onto glass slides of 0.17 mm thickness and 10 mm diameter.

**Sample characterization**: An *Olympus IX71* microscope, equipped with 0.8NA air (*Olympus LMPlanFL-100x*) and 1.4NA oil

(*Olympus UPLSAPO100XO*) objectives, was used to observe to sample mounted on a *PI P-713 XY Piezo Scanner* nanopositioning stage. Fluorescence widefield imaging used an Hg lamp (*Olympus USH-1030L*) light filtered at 438±12 nm for illumination. Single emitters were scanned confocally with a 405 nm pulsed laser (*PicoQuant LDH series* with 50 ps temporal width and operating at 2.5 MHz), and the emission was filtered spectrally by a 473 nm long-pass filter (*Semrock 473 nm RazorEdge®*) and spatially by a 150 µm diameter pinhole. Time-resolved photoluminescence and time-correlated single photon counting data was obtained with a *PicoHarp300* photon counting module and two *Micro Photon Devices PDM series* single photon avalanche photodiodes arranged in a Hanbury-Brown and Twiss setup.

## Acknowledgements

This work was supported by DIM NanoK funding through the project PATCH and by ANR DELIGHT.

## Conflict of interests

The authors declare no conflict of interests.

## Author contributions

A.R.D. and A.M. conceptualised the experiments and investigation tools. A.R.D., N.C. and W.M. performed the experiments. M.N. and B.D. synthesized the QDs. A.R.D and A.M. analysed the data, modelled it, and wrote the paper.